\documentclass[review]{elsarticle}

\usepackage{amsmath}
\usepackage{natbib}
\usepackage{color}
\usepackage{geometry}
\usepackage{fleqn}
\usepackage{graphicx}
\usepackage{txfonts}
\usepackage{hyperref}

\usepackage{bm}
\usepackage{float}
\usepackage{amsfonts}
\usepackage[normalem]{ulem}

\title{Machine learning approaches for analyzing and enhancing molecular dynamics simulations}

\author[1]{Yihang Wang}
\ead[1]{yhwang17@terpmail.umd.edu}

\author[2]{Jo\~ao Marcelo Lamim Ribeiro}
\ead[2]{joaomarcelolamim.ribeiro@mssm.edu}

\author[3]{Pratyush Tiwary \corref{cor1}}
\cortext[cor1]{Corresponding author}
\ead[3]{ptiwary@umd.edu}

\address[1]{Biophysics Program and Institute for Physical Science and Technology, University of Maryland, College Park, MD, 20742, USA}
\address[2]{Department of Pharmacological Sciences, Icahn School of Medicine at Mount Sinai, One Gustave L. Levy Place, Box 1677, New York, NY, 10029, USA
}
\address[3]{Department of Chemistry and Biochemistry and Institute for Physical Science and Technology, University of Maryland, College Park, MD, 20742, USA}

\date{August 2019}

\begin{document}

\begin{abstract}
Molecular dynamics (MD) has become a powerful tool for studying biophysical systems, due to increasing computational power and availability of software. Although MD has made many contributions to better understanding these complex biophysical systems, there remain methodological difficulties to be surmounted. First, how to make the deluge of data generated in running even a microsecond long MD simulation human comprehensible. Second, how to efficiently sample the underlying free energy surface and kinetics. In this short perspective, we summarize machine learning based ideas that are solving both of these limitations, with a focus on their key theoretical underpinnings and remaining challenges.
\end{abstract}
\maketitle

%3-5 Research highlights Required 85 characters each including spaces etc.
%\begin{highlights}
\section*{Highlights}
\begin{enumerate}
\item Machine learning and artificial intelligence approaches have been leveraged for MD.
\item One machine learning contribution is in removing noise to make MD data human accessible.
\item Yet another contribution is in helping to enhance sampling to make MD simulations more ergodic.
\item The problem of making MD data human accessible and enhancing MD sampling requires overlapping ideas.
\end{enumerate}
%\end{highlights}

\section{Introduction}
With the ever-increasing power and availability of high-performance computing resources, coupled with the development of accurate interaction models, molecular dynamics (MD) simulations have now become an indispensable tool for the study of biophysical systems. MD has allowed us to probe, with all-atom spatial and femtosecond temporal resolution, complex processes such as the folding/unfolding of a chain of residues, association/dissociation of protein-ligand systems, protein-protein interactions and countless others. Not entirely dissimilar to many other fields in science and engineering, here as well we thus have had an explosion of data, easily reaching hundreds of gigabytes for a standard microsecond long MD simulation of a protein solvated in explicit water. This immediately leads to two pressing questions. First, what do we do with this much data -- how do we store it and how do we make sense of it? And if a microsecond long simulation of a single solvated protein can generate this much data, what would happen if we tried to simulate an entire cell for (say) the same or longer duration? Second, if one microsecond long simulation of a single protein in explicit water with classical interaction models can take a few weeks, what will it take to simulate such a system at timescales actually relevant to various biological processes such as ligand dissociation and slow conformational exchanges, namely seconds, minutes and beyond? Thus MD suffers from two, at first glance opposing but deeply connected problems - enormous amounts of data which can be difficult to analyze, and yet, the inability to generate data at the timescales that we might actually care about when it comes to interpreting or designing laboratory experiments.

\begin{figure}[ht]
\centering
\includegraphics[width=.90\linewidth]{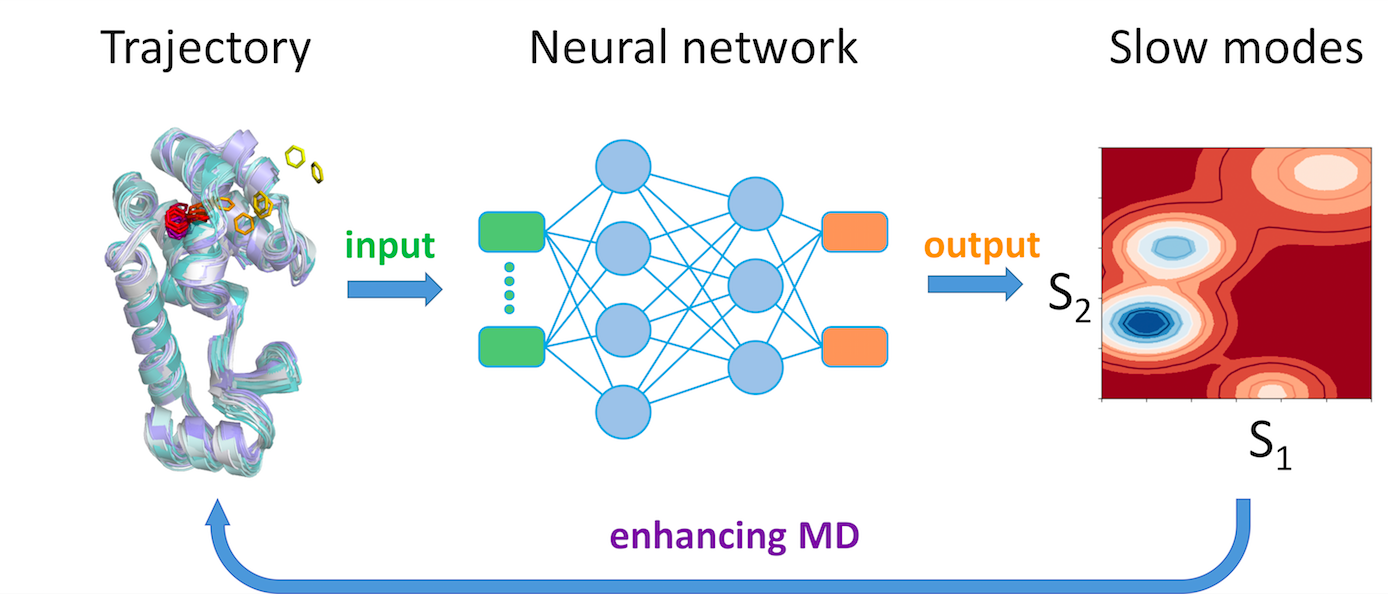}
\caption{A schematic illustrating the typical workflow of some of the methods that use machine learning to analyze and enhance MD simulation. High dimensional data which describes the time evolution of the system in configuration space is used as the input of an NN or DNN. The NN is trained to project the input to a low dimensional space. Depending on the structure of NN and objective function, the low dimensional representation captures different features that are considered to be important, such as slow modes. In some methods, the feature learnt are used to further enhanced the sampling of MD simulation as shown by the arrow in the bottom.}
\label{fig:architecture}
\end{figure}

The purpose of this review is to summarize how the last few years are starting to see significant breakthroughs in the endeavour of surmounting both these problems by using ideas from machine learning (ML), motivating how these two problems are actually two sides of the same coin. By ML here we mean methods such as neural networks (NN), deep neural networks (DNN) and related ideas that lie at the heart of the field of artificial intelligence. For the sake of completeness we also consider a few ML-like data driven approaches. Most of these methods follow some variant of the scheme shown in Fig.\ref{fig:architecture}. That is, ML is used to find a projection from the high dimensional structure space to a low dimensional feature space. Due to paucity of space we do not discuss any applications, but instead focus on the central theoretical ideas behind these approaches for analyzing and enhancing MD simulations. In addition, ML approaches to designing force-field\cite{behler2007nn,csanyi2010gaussian,roitberg2017anakin} itself will also not be covered.  We conclude with words of caution as to how in spite of its indubitable potential, ML is not a cure-all at least in the context of MD simulations, and while the future ahead is exciting, much work remains to be done.

\section{Underlying terms and constructs}
\label{sec:terms}
We begin by providing a very brief compendium of various terms and constructs relevant to the remaining parts of this review. We are interested in a generic atomic/molecular system comprising $N$ atoms, where the 3$N$ position coordinates are collectively denoted by the state vector $\mathbf{x}$. In classical MD, the $N$ atoms execute classical dynamics at a temperature $T$ (or inverse temperature $\beta$) under the influence of an interaction potential $U(\mathbf{x})$. Further, we are typically interested in cases relevant to biophysical systems characterized by the existence of many metastable states where the system spends extended amounts of time, only rarely moving between any two such states. Our objective in performing MD is then to evaluate the following two broad types of quantities for some generic low-dimensional collective variable (CV) $s(\mathbf{x})$:
\begin{enumerate}
    \item equilibrium properties, such as the free energy or potential of mean force $F(s)$, where  $ P(s) \equiv e^{-\beta F(s)} = \int d\mathbf{x} \  \delta \left[ s-s(\mathbf{x}) \right ] P_0(\mathbf{x})$, where $P_0(\mathbf{x}) \propto e^{-\beta U(\mathbf{x})}$ is the equilibrium probability of a microstate $\mathbf{x}$, and  $ P(s)$ is the probability of the CV.
    \item dynamic properties, such as the mean first passage time for escaping any metastable state in $\mathbf{x}-$space.
\end{enumerate}

 The dynamics in the full high-dimensional $\mathbf{x}$-space can be safely assumed to be Markovian, and thus the time-propagation of the system in $\mathbf{x}$-space can be described for a given lag time $\tau$ using a transfer operator\cite{klus2018data} $K$ with eigenvalues $1=\lambda_0 > \lambda_1 \geq  ...$ and corresponding eigenvectors $\psi_0(\mathbf{x}) =1,\psi_1(\mathbf{x}),\psi_2(\mathbf{x}),...$. The eigenvectors with indices $i \geq 1$ correspond to the slow modes of the system with corresponding timescales $t_i = -{\frac{\tau}{ \text{log }\lambda_i}}$.  A central objective of the methods in this review is to learn the so-called low-dimensional \textit{reaction coordinate} (RC) for the high-dimensional dynamics. Loosely speaking, this RC can be defined as a special type of CV which encapsulates sufficiently many slow modes of the high-dimensional transfer operator, so that the modes not captured are either fast or irrelevant to the process of interest.

\section{Using ML for analyzing MD trajectories}
\label{sec:analyzing}
ML has been used in a variety of forms to analyze a given long MD trajectory and learn relevant dominant processes or slow modes. Ma and Dinner as early as in 2005 used NNs for constructing a RC\cite{Dinner_automatic_identify_rc}. The idea was to sample many configurations along pathways going from one metastable state to another, and tabulate an ensemble of committor values, or the probability of committing to one of the metastable states after a given time. Given this database of values, a NN was trained to determine how the committor depended on different CVs, and identify the most important few CVs, which were sufficient for the NN to give best fit to tabulated committor values. This framework was recently revived and somewhat generalized by Hummer and co-workers \cite{jung2019artificial} through the use of deeper layers and using the learned committor to generate a more accurate ensemble of reactive pathways.

While a committor might be arguably the most rigorous way to describe a slow mode, there are many other reasonably accurate and computationally much cheaper principles defining what constitutes slow modes. One such formalism, variational approach to conformation dynamics (VAC), is based on transfer operator theory\cite{klus2018data}. VAC attempts to calculate the eigenvalues and eigenvectors of the transfer operator defined in Sec. \ref{sec:terms} starting with $\psi_1$, by solving a variational principle which essentially states that any trial eigenvector $\psi_1'$ will have a time-lagged autocorrelation $\langle \psi_1'|K|\psi_1'\rangle \leq 1$, as long as it is orthonormal to $\psi_0$, with equality holding iff $\psi_1'=\psi_1$. Thus we can approximate the first nontrivial eigenfunction $\psi_1$ by searching for a $\psi_1'$ that maximizes its autocorrelation function subject to the orthonormal conditions. Ref. \cite{noe2013variational,nuske2014variational} show how to calculate such matrix elements given unbiased MD data.  Once the first slow mode has been learned, further modes can be learned in a similar manner but with more orthonormality conditions. Methods such as TICA \cite{perez2013identification} implement VAC directly by learning slow eigenvectors as linear combinations of pre-selected basis functions, and are at the heart of building MSMs. The Variational approach for Markov processes (VAMP) principle \cite{VAMPnet} generalizes the mathematical framework in the VAC principle to nonstationary and nonreversible processes, while keeping the same key underlying intent of learning a transformation of $\mathbf{x}$ in which the dynamics is as Markovian as possible. VAMPnets use DNNs to implement the VAMP principle and automate the various steps involved in construction of a MSM.
Even more recent methods such as state-free reversible VAMPnets (SRV) combine strengths of VAC and VAMPnets for equilibrium systems, and use DNNs to construct a full hierarchy of slow modes expressible as non-linear functions of input coordinates.

Slow modes $s$ as per VAC may also be equivalently defined as a mapping $s(\mathbf{x})$ that maximizes autocorrelation $A(s)$ for a lag time $\tau$ :
\begin{equation}
A(s) = { \mathbb{E}\left[\tilde{s}(\mathbf{x}_{t}) \tilde{s}(\mathbf{x}_{t+\tau}) \right]
\over \sigma(s(\mathbf{x}_{t})) \sigma(s(\mathbf{x}_{t+\tau}))}
    \label{eq:acf}
\end{equation}
where $\tilde{s} = s -\mathbb{E}(s)$ is the mean-free latent variable and $\sigma(s) $ is the standard deviation of $s$. Time-lagged autoencoders (TAE) use an encoder-decoder framework to find the slow component so defined. However, instead of calculating the precise expectation value in Eq. \ref{eq:acf}, TAE approximates the slow mode by minimizing the reconstruction loss $\mathcal{L}_{\text{R}}$:
    \begin{equation}
        \mathcal{L}_{\text{R}} \equiv\mathcal{L}_{\text{TAE}} =\sum_{t}\left\|\mathbf{x}_{t+\tau}-D\left(E\left(\mathbf{x}_{t}\right)\right)\right\|^{2}
          \label{eq:tae}
    \end{equation}
 Here $E$ is the encoder which maps the input configuration $\mathbf{x}$ to low-dimensional $s$ and $D$ is the decoder mapping it back to coordinate space. Maximizing Eq. \ref{eq:acf} and minimizing Eq. \ref{eq:tae} are identical if the encoder and decoder are linear \cite{limit_TAE}. TAE implements both the encoder and decoder using DNNs in which case Eq. \ref{eq:tae} is an approximation to Eq. \ref{eq:acf}.

Variational Dynamic encoder(VDE) \cite{vde} also uses an encoder--decoder pair with time-lagged data, but instead of having only a term related to reconstruction error $\mathcal{L}_{\text{R}}$, VDE includes two additional terms. For the first additional term, the latent variable $s_t$ is not directly used as an input of the decoder to predict $\mathbf{x}_{t+\tau}$. Instead, a probabilistic framework is employed, where a sample $s'_t$ is drawn from a Gaussian distribution $\mathcal{N}(\mu(\mathbf{x}_{t}),\sigma^2(\mathbf{x}_{t}))$, with mean $\mu(\mathbf{x}_{t}) = E(\mathbf{x}_{t})$ as in TAE and the variance $\sigma^2(\mathbf{x}_{t})$ learnt through another NN. This forces the decoder to be tolerant to small variances of signals from latent space, thus increasing model generalizability. The use of a Gaussian prior also helps the distribution of $s'_t$ to be smooth, allowing meaningful interpolation between states in latent space. This is done by introducing the Kullback--Leibler divergence loss: 
    \begin{equation}
    \mathcal{L}_{\text{KL}}=\mathbb{E} \left[\frac{1+\log \sigma (\mathbf{x}_t)^2 -\mu(\mathbf{x}_t)^2 -\sigma(\mathbf{x}_t)^2}{2} \right]
    \label{eq:kl_loss}
    \end{equation}
    %Minimizing $\mathcal{L}_{\text{KL}}$ will force the distribution of $s'_t$ to be close to an Guassian prior. These together is referred as variational bayesian inference, which is widely used in ML to learn a generative model. 
Secondly, another term called autocorrelation loss is introduced. Autocorrelation loss $\mathcal{L}_{\text{AC}}= -A(s)$ is the negative of the autocorrelation defined by Eq. \ref{eq:acf}. It encourages the learning of modes with high autocorrelation and makes the training process easier to converge. The VDE objective function is then a sum of these three loss terms. Recent work has included a probabilistic framework similar in spirit to VDE but within a full Bayesian approach \cite{deep_Bayesia_models}.

The EncoderMap approach of Lemke and Peter is another method that makes use of a NN encoder-decoder architecture \cite{encoderMap}. In addition to a reconstruction loss analogous to Eq. (\ref{eq:tae}) measuring the distance between input configurations $\mathbf{x}_{t}$ and their reconstructions $D\left(E\left(\mathbf{x}_{t}\right)\right)$, the NN loss function used for training includes an additional term meant to force $s$ to be interpretable. This additional term is the sketch-map cost function \cite{ceriotti2011simplifying}, which aims to ``focus" the network's low-dimensional central bottleneck on learning a CV capable of reproducing distances between adjacent metastable basins as opposed to intra-basin and basins separated from each other. In contrast to sketch-maps, which are expensive to train, EncoderMap can handle large datasets since modern NNs implementations are designed for the data-rich regime.
%Since the EncoderMap loss function is written as a linear combination of these two terms (as well as a regularization term to prevent NN overfitting), it is possible to shift the emphasis from learning a CV that prioritizes the reconstruction of the molecular configurations versus constraining the distances between all available configurations to be within a certain threshold. 

A recent ML approach \cite{XGBOOST} has been proposed that does not leverage NNs like most of the above approaches. Instead, it uses the XGBoost algorithm to determine a set of essential internal coordinates. In the XGBoost algorithm, an ensemble of decision trees is generated with the aim of modeling classification rules for assigning an outcome to a given molecular configuration $\mathbf{x}$. For our purpose, the outcomes are the available metastable states and the XGBoost algorithm is a useful supervised ML approach for learning which of the input coordinates are most useful to perform accurate classification of configurations into the different metastable states, with the most useful coordinates in $\mathbf{x}$ being those that lead to the largest gains in the loss function \cite{XGBOOST,chen2016xgboost}. However, it is possible for artifacts to appear in the ranking of these coordinates, since in general these coordinates can be highly correlated to one another. In this work the authors thus propose -- in addition to training a classification model with XGBoost \cite{chen2016xgboost} that ranks the coordinates -- a loop that removes the highest ranking coordinate as per XGBoost and re-trains a new classification model \cite{XGBOOST}. This allows the importance of the different MD features to be determined ``in isolation" from other highly important coordinates \cite{XGBOOST}.

In recent work \cite{Dynamic_graphica_models}, Olsson and No{\'e} have extended the notion of encoding a global molecular configuration, $\mathbf{x}$, into encoding several local configurations $\mathbf{x}^1$, $\mathbf{x}^2$,..., $\mathbf{x}^j$, each representing a partition of the original global molecular structure into a local substructure. To model the time evolution of $\mathbf{x}$ the propagator or conditional distribution $p(\mathbf{x}_t|\mathbf{x}_{t-\tau})$ is written in terms of the substructures:
\begin{equation}
p(\mathbf{x}_t|\mathbf{x}_{t-\tau}) \propto e^{\sum_{i} \mathbf{x}^i_{t}(\sum_{j} J_{ij}(\tau) \mathbf{x}^{j}_{t-\tau} + h_{i}(\tau))} 
\end{equation}
where $J_{ij}$ is the coupling parameter between the \textit{i$^{th}$} and \textit{j$^{th}$} subsystem and $h_{i}$ describes the coupling between the \textit{i$^{th}$} subsytem with an external field. With the choice of this model, the problem of determining the coupling parameters can then be reduced to $N$ logistic regression problems. A notable feature of this approach is that it seems capable of predicting molecular configurations that have not been incorporated into the training data. Prediction of unsampled but likely configurations along with a quantitative measure of their relative likelihoods is exactly the theme of our next section.

\section{Using ML and related data-driven approaches to enhance sampling}
\label{sec:enhancing}
In this section, we review approaches that use ML to not just analyze existing MD generated structures and trajectories, but also actively enhance the sampling capacity of MD. In other words, these approaches use ML to not just learn from given data, but actually generate statistically accurate information when the underlying processes are so slow that they can simply not be sampled in unbiased MD even with the best available computing resources. Generating novel low-probability structures is not hard -- simply heat the system. What is extremely non-trivial is generating them so that their statistics, typically through some corrective reweighting scheme, is in accordance with the underlying Boltzmann probability, and even more ambitiously, with the underlying kinetics at least in terms of inter-conversion timescales between different metastable states. This is a complex problem for which many exciting solutions have recently been proposed, and here we outline some of them. Essentially, the success of ML highly relies on the abundance of data. However if the events of interest are rare, one faces paucity of relevant data to train the ML upon. One could train ML models on data from enhanced sampling methods, but these methods themselves need an estimate of the RC, and a wrong choice of RC used to generate the data could mislead the ML being trained upon it. One solution to this dilemma is to iterate between rounds of sampling and ML, where every ML round generates a progressively improved RC, which is used to perform better enhanced sampling and so on. Convergence of the RC and associated information between iterative rounds of sampling and learning can be viewed as a necessary though not sufficient condition for the reliability of such a protocol, though many interesting mathematical questions arise regarding why such a procedure should converge to the true RC and associated thermodynamic/kinetic observables. Many (though not all) of the schemes described in this section are built around this central idea, differing in the precise forms of (a) the ML procedure, (b) enhanced sampling scheme and (c) how exactly information from ML is used to perform sampling and vice-versa.

The molecular enhanced sampling with autoencoders (MESA) approach uses a DNN with non-linear encoder and decoder to learn the RC from input data, which itself is generated through umbrella sampling along trial RCs \cite{autoencoders_Ferguson}. Every round of ML leads to an improved RC along which new umbrella sampling is performed. The iterations are continued until the free energy from umbrella sampling no longer varies with further iterations. Similar to MESA, nonlinear RCs learned by methods like VDE \cite{vde} can also be used to perform enhanced sampling, typically using TICA modes as input variables. 
%Also, being inspired by transferable learning, it is proposed that if we can use the RC learned by VDE from one system to enhance the sampling of another similar system. To do this, a mapping between the OPs of two systems needs to be learnt. In some cases, this can be done easily. For example, if alpha carbon contact distances for a wild type protein is used to learned a RC, then a simple sequence alignment can be use to find the corresponding new alpha carbon contact distances for a mutant.

Reweighted autoencoded variational Bayes for enhanced sampling (RAVE) \cite{rave, prave} is an iterative ML-MD method motivated by the observation that many feature learning methods, in addition to classifying features, also provide the probability distribution in feature space \cite{wetzel2017unsupervised}. The learnt features and their probability distribution can then respectively be used as RC and its fixed or static bias can then be applied to $U(\textbf{x})$ leading to more ergodic exploration. One crucial distinguishing feature of RAVE is that it avoids additional biasing along the RC as in umbrella sampling \cite{Umbrella_sampling,mezei1987adaptive} or metadynamics \cite{metad_ARPC}, as these could forcefully lead to enhanced sampling through non-equilibrium ways. In other words, even if the RC from ML was very far from the truth, when used in umbrella sampling especially with the typical post-processing WHAM protocol, one could still obtain some sort of free energy profile and have no way to tell how erroneous the RC from ML was. On the other hand, a static bias would lead to enhanced exploration only if the orthogonal hidden modes are not relevant. This serves as a test in RAVE that helps weed out spurious local minima solutions that often plague deep learning. Keeping the transition states between different features devoid of bias also allows obtaining pathways and rate constants.  To learn these features and their probabilities, RAVE uses a past-future information bottleneck optimization scheme that outputs a minimally complex yet maximally predictive model. A DNN decoder is trained to predict the future state of the system instead of only trying to recover the input data, and a linear encoder is used to get an interpretable projection from the space of order parameters to the RC.

Instead of using enhanced sampling methods to explore new possible configurations, deep generative markov state models (DeepGenMSM) \cite{Deep_Generative_MSM} use a generative NN to predict the future evolution of the system and thus propose new configurations. The encoder has a SoftMax layer which gives the probability of mapping an input configuration to different discretized states in latent space. A generative model is fit to predict the time-delayed evolution of the system by minimizing a suitably defined energy distance, which measures the difference between the transition density of the system and that of the generative model. The generative model is then essentially extrapolated to produce high dimensional structures that were not in the training database.

%By optimizing the NN, states on input configuration space is clustered according to its likelihood of mapping to different clusters given by the encoder and the generative model is trained to give prediction of the evolution of the system given the knowledge that which cluster the present state of system belongs to.   

% Interesting paper. It uses flow-based methods for performing inference, which is an alternative to variational inference. After training, the NN has learnt a reversible mapping between original input and a coordinate representation under the condition that in this new coordinate the distribution is simple and we can sample from it. The real crucial theoretical thing that the authors' did, is to use i) fact that mapping is one-to-one/invertible so that there is an expression/relation between probabilities in the different representations and ii) fact that functional form of target marginal distribution has a known form. i) + ii) allow for a straightforward derivation of a loss term that does not require "training by reconstruction". I do not think, however, that the method learns a slow mode since the determinant that the appears in flow-based inference means the mapping is to a space of dimension equal to the input (i.e. there is no dimensionality reduction).  
Boltzmann Generators are a very recent deep learning based approach that learns the equilibrium probability $P_{0}(\mathbf{x})$ without resorting to running long trajectories\cite{Boltzmann_Generators}. It leverages recent advances in probabilistic generative modeling in which invertible coordinate transformations mapping $\mathbf{x}$ onto a random variable $\mathbf{x'}$ whose distribution is straightforward to sample are learnt\cite{flow_method2016,flow_method2018}. Using $\mathbf{x'}$ together with the fact that $\mathbf{x}$ follows the Boltzmann distribution, $P_{0}(\mathbf{x}) \propto e^{-\beta U(\mathbf{x})}$, the NN can be trained (in principle) without using maximum likelihood on a pre-existing dataset. This is in contrast to common applications of NNs such as, for instance, image generation, where a large dataset is available but the functional form of their distribution is not known a priori. Starting with the KL divergence of the probability predicted by the NN relative to the exact (and simple) distribution in $\mathbf{x'}$-space, the loss function becomes
\begin{equation}
\label{boltz_gen_loss1}
\mathbb{E}\left[\beta U(\mathbf{x}) - log |J(\mathbf{x'})|\right]
\end{equation}
where the expectation is calculated with respect to samples drawn from the exact distribution in $\mathbf{x'}$-space, $\mathbf{x}$ is the output of the generative network that describes the inverse mapping from $\mathbf{x'}$ back to $\mathbf{x}$, and the Jacobian $J$ describes the effect of the coordinate transformation on the distribution. Notice that lowering the loss function given in Eq. (\ref{boltz_gen_loss1}) tends to lead the network towards learning a transformation that results in sampling low-energy configurations (i.e. approach the Boltzmann distribution). In practice, however, Boltzmann generators do also use maximum likelihood on a pre-exisiting dataset in order force the network to give non-negligible probabilities to other metastable states in addition to the global minima. The NN based variationally enhanced sampling (VES) method also stands out with respect to many of the other ML methods mentioned here as it does not try to learn slow modes, but instead tries to express the bias as a smoothly differentiable DNN potential as a function of pre-selected small number of CVs. To learn such a bias, it optimizes the objective function introduced by Valsson and Parrinello \cite{valsson2014variational} which has several elegant properties such as convexity.
%In a sense then, this method is similar in spirit to the methods in section ?? that use NNs to learn better approximations of the free energy, as the perfect bias in VES is an adequately scaled version of the underlying free energy.

We now mention some data driven approaches which while not using ML in its strictest definition, are connected with ML type ideas. We begin with the diffusion map directed MD (DM-d-MD) \cite{preto2014fast} and variants thereof, which are a series of pioneering methods that use non-linear manifold learning techniques such as diffusion maps \cite{coifman2005geometric} to gradually build up the slow modes of a system. The central idea in these methods is to start with a short unbiased MD run, perform diffusion map (or a similar method) on it, and then use this diffusion map to select coordinates in configuration space for launching new rounds of unbiased MD (unbiased apart from the implicit bias in selection of initiation points). The key differences in various flavors of such an approach arise in how the new launch-off points are selected. In the extended DM-d-MD approach for instance, the new starting points are picked uniformly along the two slowest timescales of the diffusion map. In the more recent iMapD approach\cite{Intrinsic_map}, the boundary in a low-dimensional manifold comprising the top few eigenvectors of the diffusion map is constructed, and independent simulations are launched from points sampled from this boundary. Both these approaches lead to enhanced exploration, however it is extremely non-trivial to obtain a Boltzmann-weighted ensemble directly from such approaches. The extended DM-d-MD method in principle alleviates this problem by assigning weights to different trajectories as they are launched so as to keep an overall Boltzmann weight. However doing this systematically and accurately might lead to drop in the computational speed-up relative to unbiased MD that one originally sought.
 
The active enhanced sampling (AES) approach from Ref. \cite{zhang2018unfolding} is similar to RAVE in the sense of iterating between enhanced sampling along a trial slow mode, and using the sampling to improve the slow mode definition. The enhanced sampling is carried out using well-tempered metadynamics \cite{metad_ARPC}. The slow mode learning is carried out using a stochastic kinetic embedding formalism that aims to learn a low-dimensional projection that is kinetically as truthful as possible to the underlying high-dimensional reality by assuming an implicit diffusion process. Each progressive round of metadynamics is carried out on top of a bias along the improved slow modes, where this bias is constructed through the reweighting formalism of Ref \cite{t_independent_fe_metad}. 

SGOOP \cite{sgoop, multi_sgoop} is an iterative method similar to RAVE and MESA that uses rounds of enhanced sampling to learn a progressively improved RC. However to learn the RC, instead of ML, a maximum path entropy or Caliber model \cite{dixit2015inferring} is learnt that identifies the RC as the low-dimensional projection with maximum separation of timescales between visible and hidden modes.

Finally, the reinforcement learning based adaptive sampling (REAP) method of Ref. \cite{shamsi2018reinforcement} learns relevant CVs on-the-fly as exploration of the landscape is carried out. REAP, like almost all other methods here, also starts with a dictionary of OPs $s$, with a starting weight $w_i$ associated with each OP $s_i$. A round of unbiased MD is carried out, then clustered into states, after which the weights $w$ are adjusted in order to maximize a reward function. In a nutshell, the reward function is designed to favor the least populated clusters. Structures from clusters with highest reward are used to then initiate a new round of MD simulations. While there is no guarantee that REAP leads to Boltzmann-weighted sampling directly or through some reweighting procedure, it does enhance the sampling and overcome the issue of orthogonal barriers.

\section{Software}
While the development of new algorithms is important and thus was the focus of this review, it is equally important to have efficient software implementing these algorithms in an accurate manner. Thankfully there is no dearth of such software. PYEMMA, PLUMED and ANNCOLVAR, as well as associated modules and scripts provided in GitHub repositories of various publications\cite{pyemma,plumed_nature,Anncolvar}, make it possible to implement many of the algorithms listed in this review.

\section{Conclusions}
In this overview we have summarized some recent ML-based methods  for analyzing and enhancing MD simulations. This is a very lively field with multiple approaches published even during the course of this review being written. These approaches are making it possible to compress high dimensional data generated during MD into low dimensional models, arguably in a more robust and automatic manner than achievable with previous non-ML methods, and revealing hidden patterns that might not have been discernible otherwise. Thus while there is clear progress, we would argue that the field is still full of several difficult, exciting open questions. First, what did the ML model learn, or the interpretability challenge. Second, would the ML model still work for small (or large) perturbations in the system being studied, or the transferability challenge. Thirdly, can a fitted ML model be used to generate Boltzmann-weighted samples that were previously unexplored, or the sampling challenge. These challenges are not very different from those faced by ML in application domains outside biology, and thus a lot is to be gained from cross-pollination between ideas from active ML experts across different domains. We thus conclude with cautious optimism for the future. \newline

\section{Acknowledgments}
The authors thank Deepthought2, MARCC, and XSEDE (Projects CHE180007P and CHE180027P) for helping our group with computational resources used to develop some of the methods reported in this work. Y.W. would like to thank NCI-UMD Partnership for Integrative Cancer Research for financial support.

\section{Highlighted References}
$\bullet$ Ref. \cite{klus2018data} -- Transfer operator formalism

$\bullet$ Ref. \cite{VAMPnet} -- Automatic MSM generation by combining the VAMP principle with NNs.

$\bullet$ Ref. \cite{vde} -- Encoder-decoder NN framework for learning the RC. 

$\bullet$ Ref. \cite{prave} -- Encoder-decoder NN framework for learning the RC as well as a static bias.

\vspace{1in}
%\textbf{References}
\bibliographystyle{unsrt}

\end{document}